\documentclass[9pt,twocolumn,twoside]{osajnl}

\journal{arxiv}

\setboolean{shortarticle}{False}

\title{Silicon microring synapses enable photonic deep learning beyond 9-bit precision}

\author[1,*]{Weipeng Zhang}
\author[1,2]{Chaoran Huang}
\author[1]{Hsuan-Tung Peng}
\author[1]{Simon Bilodeau}
\author[1]{Aashu Jha}
\author[1]{Eric Blow}
\author[1,5]{Thomas Ferreira de Lima}
\author[3,4]{Bhavin J. Shastri}
\author[1]{Paul Prucnal}

\affil[1]{Department of Electrical and Computer Engineering, Princeton University, Princeton, NJ 08540, USA}
\affil[2]{The Chinese University of Hong Kong, Shatin, N.T., Hong Kong SAR, China}
\affil[3]{Department of Physics, Engineering Physics and Astronomy, Queen’s University, Kingston, Ontario K7L 3N6, Canada}
\affil[4]{Vector Institute, Toronto, Ontario M5G 1M1, Canada}
\affil[5]{Now at NEC Laboratories America, Princeton, NJ 08540, USA}
\affil[*]{Corresponding author: weipengz@princeton.edu}

\dates{Compiled \today}

\doi{\url{http://dx.doi.org/xx.xxxx/XX.XX.XXXXXX}}

\begin{abstract}
Deep neural networks (DNN) consist of layers of neurons interconnected by synaptic weights. A high bit-precision in weights is generally required to guarantee high accuracy in many applications. Minimizing error accumulation between layers is also essential when building large-scale networks. Recent demonstrations of photonic neural networks are limited in bit-precision due to crosstalk and the high sensitivity of optical components (e.g., resonators). Here, we experimentally demonstrate a record-high precision of 9 bits with a dithering control scheme for photonic synapses. We then numerically simulated the impact with increased synaptic precision on a wireless signal classification application. This work could help realize the potential of photonic neural networks for many practical, real-world tasks. ©2022 Optica Publishing Group. Users may use, reuse, and build upon the article, or use the article for text or data mining, so long as such uses are for non-commercial purposes and appropriate attribution is maintained. All other rights are reserved.
\end{abstract}

\setboolean{displaycopyright}{true}

\begin{document}

\maketitle

\section{Introduction}
\begin{figure*}[ht!]
\centering
\includegraphics[width=.81\textwidth]{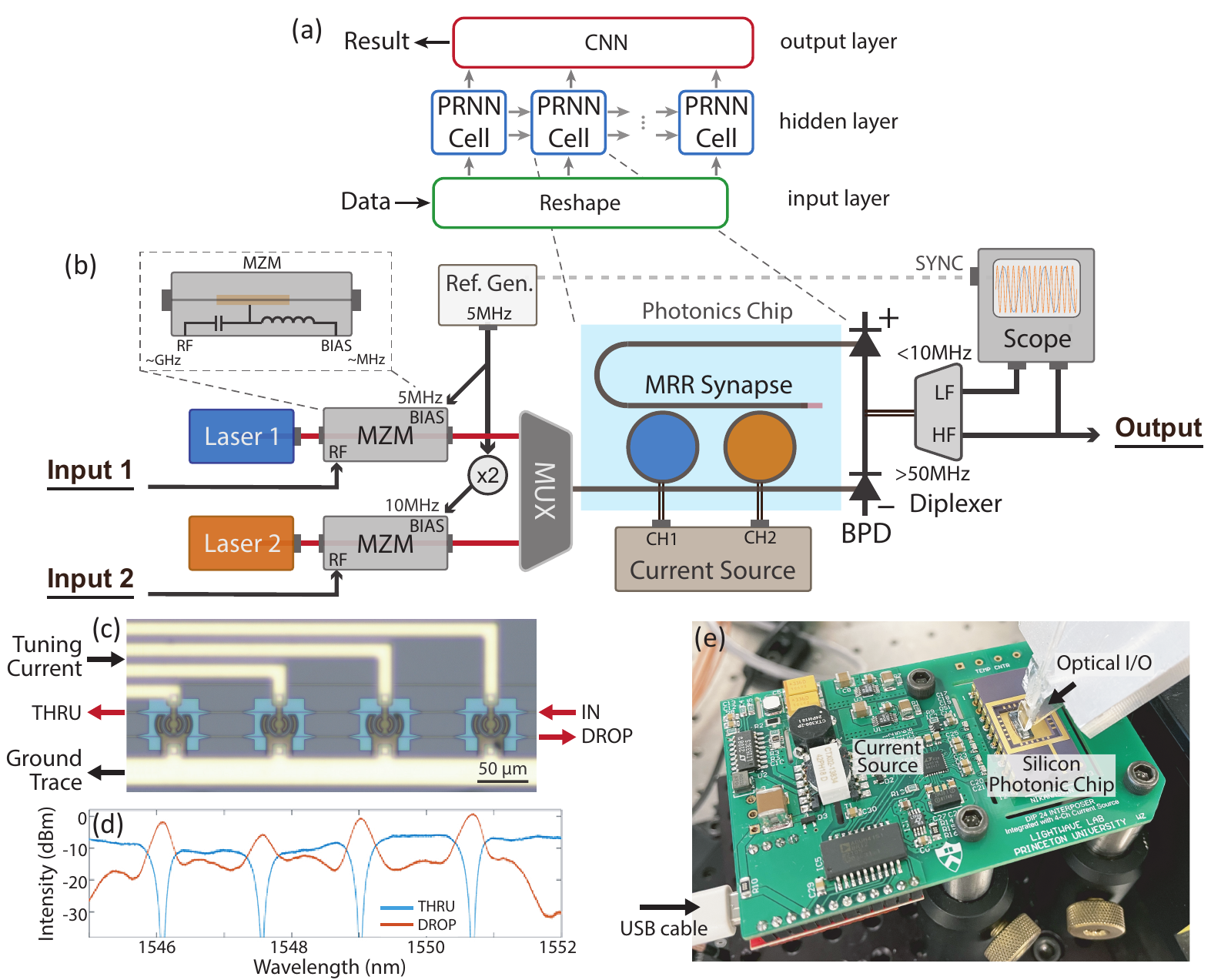}
\caption{(a) Diagram of a typical photonic neural network (the continuous-time recurrent neural network, CTRNN \cite{tait2017neuromorphic}), in which, the hidden layer is built by interconnected photonic recurrent neural network (PRNN) cells. Each PRNN cell is realized by many microring synapses similar to the one shown in (b). The input and output layers are convolutional neural networks (CNN) by electronic components. (b) Schematic of two photonic synapses with dithering control. Ref. Gen., reference signal generator. MUX, wavelength multiplexer. BPD, balanced photodetector. LF, low frequency. HF, high frequency. The inlet displays the build-in bias-tee of the MZM. (c) Zoomed-in micrograph of the MRR synapse cascading four MRRs in a parallel add/drop configuration. (d) Spectral response of the MRR synapse measured from both the THRU and DROP ports. (e) Interposer PCB. The chip, which is wire-bonded on a DIP 24 chip carrier, is mounted on this PCB. A multi-channel current source is integrated on the broad and applies currents to the MRRs. The optical input and output are through the fiber V groove on the top right.}
\label{fig:schematic}
\end{figure*}

Deep neural networks (DNN) have enabled various applications, from fundamental research in chemistry \cite{mater2019deep} and biology \cite{angermueller2016deep} to examples in daily life such as financial analysis \cite{heaton2017deep} and cybersecurity \cite{xin2018machine}. In DNNs, consecutive layers of neurons (represented as vectors) are interconnected by synaptic weights (represented as matrices). Practical implementations of DNNs often rely on certain precision in weighting to maintain usable prediction accuracy and scale-up network sizes \cite{judd2018proteus,mellempudi2019mixed}. This is especially true with noisy input signals \cite{nazare2017deep}, whose fluctuations can propagate and accumulate from one layer to the next. The state-of-the-art DNNs implemented in electronics, like the TPU \cite{jouppi2017datacenter}, have a precision of 8 bits or higher. Emerging technologies such as photonic or optical neural networks (PNN or ONN) are competitive \cite{hughes2018training,feldmann2019all}, given their fundamental benefits in terms of higher interconnection density, broad bandwidth, and lower energy consumption \cite{prucnal2017neuromorphic,shastri2021photonics}. Optical synapses are elements that configure the connection strength, or weights, between two optical neurons of consecutive layers. In Figure \ref{fig:schematic}(a), which presents an instanciation of a continuous-time recurrent neural network (CTRNN), these would loosely correspond to the arrows between different photonic recurrent neural network blocks (PRNNs) \cite{prucnal2017neuromorphic}.
However, all demonstrated photonic synapses had been limited to low precision: for example, phase-change materials (PCM) based synapses have been reported to have 5 bits of precision \cite{li2019fast}; Mach-Zehnder interferometer (MZI) based coherent networks \cite{shen2017deep,demirkiran2021electro} are susceptible to fabrication error \cite{bandyopadhyay2021hardware} which poses challenges for maintaining precision when the network scales up; and photonic synapses based on microring resonators (MRR) have so far been limited to 7 bits \cite{mrrcntr20}. Low precision undermines the inherent advantages of scalability and restricts the achievable accuracy of optical neural networks, limiting their practical applications, such as the classification of wireless signals.

Resonator-based approaches to optical synapses, such as MRRs \cite{mrr12}, are appealing due to their compactness, sensitivity, and innate wavelength-division multiplexing (WDM) compatibility. However, this resonant nature results in inter-channel crosstalk (from overlapping spectral filter responses) and susceptibility to spurious environmental fluctuations (from high sensitivity). Many methods have been proposed to improve weighting control, including feedback methods \cite{mrrfb18} to address the high sensitivity and a feedforward model \cite{mrrcntr16} to overcome inter-channel crosstalk. For years, however, the control precision of MRR-based photonic synapse has seemingly plateaued at around 7 bits \cite{mrrica20, mrrcntr20}. 

The limited precision, from our perspective, is due to the lack of monitoring of the entire signal path of the photonic synapse. Previous control methods only monitor drifts of the MRR resonances and are insensitive to other fluctuations. We observed that every component in the input signal path could cause weight drifting and should be considered. For example, the electrical-to-optical modulator is the component that converts the input electrical signal onto the photonic path. If its modulation depth fluctuates (usually caused by polarization drift or temperature drift), the amplitude of the output signal, namely the weight, will drift even if the MRRs remain perfectly stable. This scenario, however, is not necessarily captured by the other feedback methods since the MRR remains on resonance. Thus, the current performance bottleneck is caused by drifting sources other than the MRR synapse. To compensate for the thermal drift of a single microring, a dithering scheme \cite{padmaraju2012thermal,padmaraju2013integrated,padmaraju2013wavelength,padmaraju2014resolving} was proposed to lock the microring to its resonance frequency in a static fashion. Photonic synapses, however, require a novel implementation of the “dithering” since more than one device needs to be controlled, and each is dynamically tuned for weighting as opposed to being statically stabilized. Also, other elements in synapses, besides the MRR itself, can induce drifts in addition to the thermal drift.

Here, we develop a new dithering scheme to control multiple photonic synapses simultaneously and experimentally achieve a record high precision of 9 bits. This scheme introduces a dithering control signal to inputs, allowing monitoring and stabilizing of the entire optical link comprising the photonic synapses, from the MZM (where the RF signals enter the system) to the BPD (where they exit the system), as shown in Figure \ref{fig:schematic}(b). With this, drifts of weighting caused by environmental changes can be dynamically tracked and compensated along the entire link. More importantly, our scheme can also track and compensate for the drift due to inter-channel crosstalk. Thus, accurate weighting can be guaranteed, especially for high fan-in neurons with multi-channel frontends (multiple MRRs). We tested this method on a two-MRR synaptic system (i.e., a neuron with inputs) and achieved about 9 bits of precision, which, to our knowledge, is the highest recorded bit precision in a photonic neuromorphic architecture. We also simulated a photonic neural network for a real-world wireless signal classification problem, and the result shows that increasing the weighting precision from 5 bits\cite{mrrfb18} to 9 bits will increase the classification accuracy from only 50\% to over 90\% in some cases. Such an improvement makes photonics competitive against its analog electronic counterparts, paving the way for practical photonic neural networks for real-world tasks, including but not limited to intelligent signal processing of RF signals \cite{marpaung2019integrated,mrrbss20} and optical communications \cite{huang2020demonstration}.

\begin{figure}
\centering
\includegraphics[width=.95\linewidth]{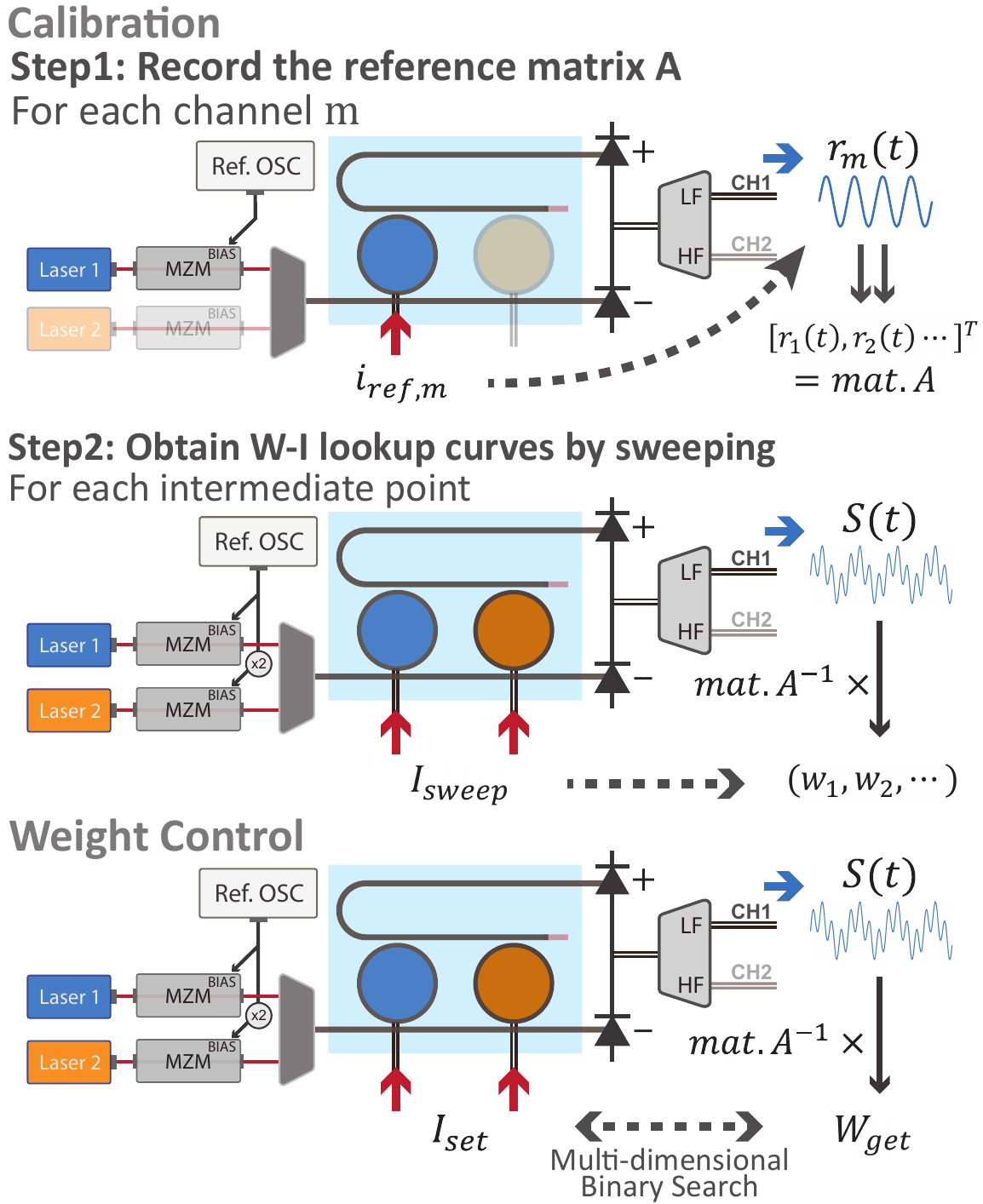}
\caption{Procedure of the dithering control. Mat. matrix. The calibration procedure consists of two steps. Step 1: Set current to zero and record the dithering waveforms as the reference of weights equal to one. Step 2: Sweep the current of all the MRRs, record the dithering waveforms and calculate the weights for each MRR. This step generates a weight-current lookup curve for each MRR. The weight control procedure performs a multi-channel binary search to tune the MRRs to the commanded weights. The lookup curves obtained in the calibration procedure provide the initial searching points for this weight control procedure. 
}
\label{fig:flowchart}
\end{figure}

\section{Method and Setup}
MRRs are WDM-compatible tunable filters where multiple MRRs of different radii can provide parallel weighting to lightwaves of different wavelengths. Accordingly, MRR-based photonic synapses can be realized through the "Broadcast-and-weight" architecture \cite{tait2014broadcast}. This architecture establishes connections between a pair of neurons where input signals from each upstream neuron are amplitude-modulated at different carrier wavelengths, then filtered or "weighted" by MRRs, and summed optoelectronically, resulting in an optical output signal that is sent to downstream neurons \cite{shastri2021photonics}. In this way, the weighting precision of each synapse is key to the overall performance of the photonic neural network. 

A typical MRR-based synaptic system is depicted in Figure \ref{fig:schematic}(b), where the signal path starts at the Mach-Zehnder modulators (MZM) and ends at the photodetector. The optical intensity delivered from the laser sources to the MRR weights is constant in an ideal system. In practice, however, polarization and temperature drifts, unstable optical alignment, or changing electrical parasitics can cause uncompensated power fluctuation resulting in an effective weight error. Fortunately, our dithering control scheme can address all these non-idealities by monitoring the entire signal path, bringing about performance improvement of photonic NNs.

As shown in Figure \ref{fig:schematic}(b), we use an MRR weight bank that can process two signal channels (see detailed description in Ref. \cite{mrrbss20}). To enable dithering control in the system, we superimpose a dithering signal, a predefined sinusoidal wave (generated by 33220A, Keysight) with a frequency usually much lower (< 10MHz) and an amplitude much smaller (< 100mV) than the input signals, into each signal path. These dithering signals can be separated with a diplexer (ZDPLX-2150-S+, Mini-Circuits) and captured synchronously by the oscilloscope (DPO4032, Tektronix). The oscilloscope is triggered by the sync output of the reference signal generator to align each captured waveform of the dithering signals. Since dithering signals share the same path with the signals being weighted, the variations of the output dithering signals reflect the actual weight value in real-time. Thus, an accurate weight can be continuously monitored regardless of the input data statistics, such as sparsity, variance, etc.

The dithering signal is introduced through the built-in bias tee of the MZMs (see the inset of Figure \ref{fig:schematic}(b)). Instead of applying pure DC voltages to the bias ports, we dither the bias voltages at predefined frequencies while the RF input port remained unchanged. Thus, the only additional components required are a diplexer and a low-end oscillator. Therefore, this method can be seamlessly integrated into many demonstrated optical synaptic setups. Scaling this dithering technique can be done in two ways. For a small system, say less than four microrings in total, new dithering signals of other harmonic frequencies must be added, which can be achieved using a frequency doubler or new generators. For a larger system (> 4 microrings), a more affordable and efficient way is to use an RF switch. In this way, the dithering signal can be applied to one microring at a time in a time-multiplexed manner. Both scenarios require only one diplexer and one scope channel to separate and record the dithering signals, which do not scale with the number of synapses and help maintain a low power budget. We demonstrate the first strategy using a 2 MRR synapse system. This method eliminates the need to directly sense the MRR resonances locally via a dedicated circuit, as with the feedback methods. We can replace the source-measure-units (SMU) with voltages/current sources for driving MRRs, usually of lower cost and smaller footprint, as shown in Figure \ref{fig:schematic}(e). Removing sensing also eases the electrical overhead required for large-scale MRR synapse actuation. We include more details of the PCB design in the supplemental material.

The operational principle of this dithering control method is illustrated in Figure \ref{fig:flowchart}. First, a two-step calibration process is done to obtain a weight-current ($W\times I$) lookup curve for each channel, providing initial searching points for the subsequent weight control process. During calibration, the first step (refer to step 1 in Figure \ref{fig:flowchart}) is to tune all the MRRs to a reference position by applying predefined fixed currents ($i_\textrm{ref,m}$). Meanwhile, the MZMs are sequentially dithered, and the output waveforms (captured by the scope) compose the reference dithered signals ($r_m$, for each channel $m$). Given the sampling length of $N$ and the total number of channels $M$, the recorded reference signals can be expressed as $r_m (t)$, where $t=0,\Delta T,2\Delta T,3\Delta T,\ldots,(N-1)\Delta T$ and $m=1,2,\ldots, M$ (index of the channel). Then a reference matrix A can be constructed using all the $r_m (t)$ as its column (as shown in Eq. \ref{eq:1}).
\begin{equation}
\begin{split}
A & = [r_1(t), r_2(t), \cdots , r_M(t)]^T\\
& =
\begin{bmatrix}
r_1(0) & r_1(\Delta T) & \cdots & r_1((N-1)\Delta T)\\
r_2(0) & r_2(\Delta T) & \cdots & r_2((N-1)\Delta T)\\
\cdots & \cdots & \cdots & \cdots\\
r_M(0) & r_M(\Delta T) & \cdots & r_M((N-1)\Delta T)\\
\end{bmatrix}
\end{split}
\label{eq:1}
\end{equation}

The next step (step 2 in Figure \ref{fig:flowchart}) is to turn dithering on for all the channels and do a full-range multidimensional current sweep of all channels. The scope records frames of the dithering signals at each intermediate point in the sweep. For this procedure, each captured waveform, $S(t)=(s(0),s(\Delta T),\ldots,s((N-1)\Delta T))$ represents the addition of all the dithering signals. The amplitude of each dithering frequency can be decomposed using Eq. \ref{eq:2},
\begin{equation}
W = (w_1,w_2,\cdots,w_M)
= \underbrace{A^{-1}}_{M\times N} \times \underbrace{S}_{N\times 1}
\label{eq:2}
\end{equation}

where $A^{-1}$ is the pseudo-inverse matrix of matrix $A$. In the vector $W=(w_1,w_2,\ldots,w_M)$, each element $w_m$ represents a correlation between the weighted sum waveform and the reference dithering signals, therefore corresponding to the real-time measurement of the weights. Note that this sweep only needs to be carried out once regardless of the channel count since each captured waveform can decompose the weights for all the channels. Then, we can obtain the $W\times I$ (weight-current) lookup curves corresponding to the applied current and the output weight for each channel, as step 2 of the calibration procedure shown in Figure \ref{fig:flowchart}. In our lab setup, the frequency of the calibration step needs to be done once a day. Whenever a new set of weights needs to be commanded, the dithering will be turned on, allowing the subsequent binary search to precisely tune the microrings to the target weights. The lookup curves obtained in the previous calibration procedure provide the starting points. The binary search iteratively adjusts the current setpoints ($I_{set}$) of the microring according to the measured weight outputs ($W_{get}$), as the weight control procedure shown in Figure \ref{fig:flowchart}). It is worth noting that binary searching is carried out with multi-dimensional searching, which means the weights of all the channels are measured and adjusted simultaneously without scaling search time with channel count. The multi-dimensional searching also minimizes inter-channel crosstalk since the applied currents account for inter-channel interference. More details on this weight searching process can be found in section 1 of the Supplemental Information.

\begin{figure}
\centering
\includegraphics[width=.95\linewidth]{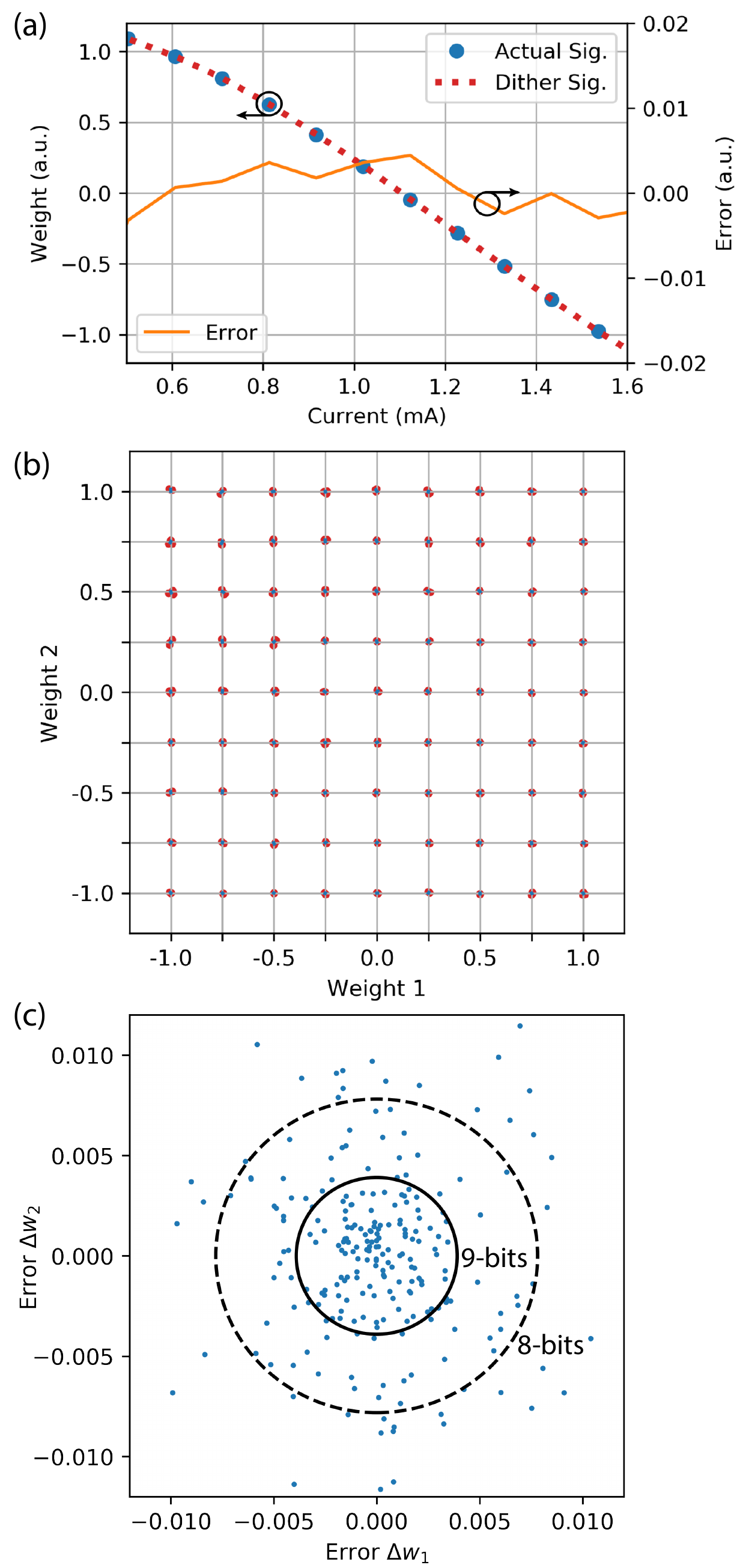}
\caption{(a) Weight comparison of the actual signal and dithering signal. (b) Measurement of weighting precision in two MRR synapses. The synapse precision is evaluated at weights on the grid. Each red dot represents a measured weight, and three measurements were performed at each point on the grid. (c) Weighting error for the trials in (b), calculated as follows ($\Delta w = w_{meas}-w_{target}$) where $w_{meas}$, $w_{target}$ are the measured and target weights respectively. The dashed and solid circles correspond to 8, and 9 bit-precisions, respectively.
}
\label{fig:meshweight}
\end{figure}

\section{Result}
To show the effectiveness of the dithering control method, we evaluated the agreement between the dithering signal and the actual signal. We apply a 10 MHz dithering signal atop a 120 MHz sine wave representing the synaptic signal. After the calibration steps, we swept the current (roughly from 0.5 to 1.6 mA) applied to the MRR heater to vary the weights in equally spaced increments within a [-1,1] interval as shown in Figure \ref{fig:meshweight}(a). Meanwhile, we recorded the weights derived from the output signal waveforms (high-frequency output of the diplexer) and the dithering signals (diplexer low-frequency output) on two oscilloscope channels. Figure \ref{fig:meshweight}(a) shows the measured weights for various applied currents. The dashed and solid curves represent the weights (left axis) measured by the actual and dithering signals. The error between the two weight values, calculated as the orange curve in the figure, is within $0.5\%$ across the whole range. This confirms that this dithering method can be used as an accurate proxy of the actual weight.

Next, the performance evaluation was extended to use two MRR synapses. Figure \ref{fig:meshweight}(b) shows the result on a mesh-plot. This type of mesh-estimation was used in previous work \cite{mrrcntr16,mrrcntr20}, in which two MRRs are tuned to equidistant grid points of $(w_1,w_2),w_{1,2}\in[-1,1]$, and the weight accuracy is evaluated at each grid point. The standard deviation of the measured errors at each point ($\sim$0.0039) gives the equivalent bit-precision of the weight control performance. We chose a 9x9 mesh, and for each grid point, we repeated the weight search three times to test the repeatability of the process. Figure \ref{fig:meshweight}(c) shows the aggregation of the errors derived from all the tested points. Compared with previous work, the error is significantly reduced, and a precision of 9.0 bits is achieved. This reflects a four-fold (or 2-bit) improvement from the previously best-reported result \cite{mrrica20}, and over 5-bit improvement from the first reported MRR synapse \cite{mrrwb16}.

\begin{figure}
\centering
\includegraphics[width=.9\linewidth]{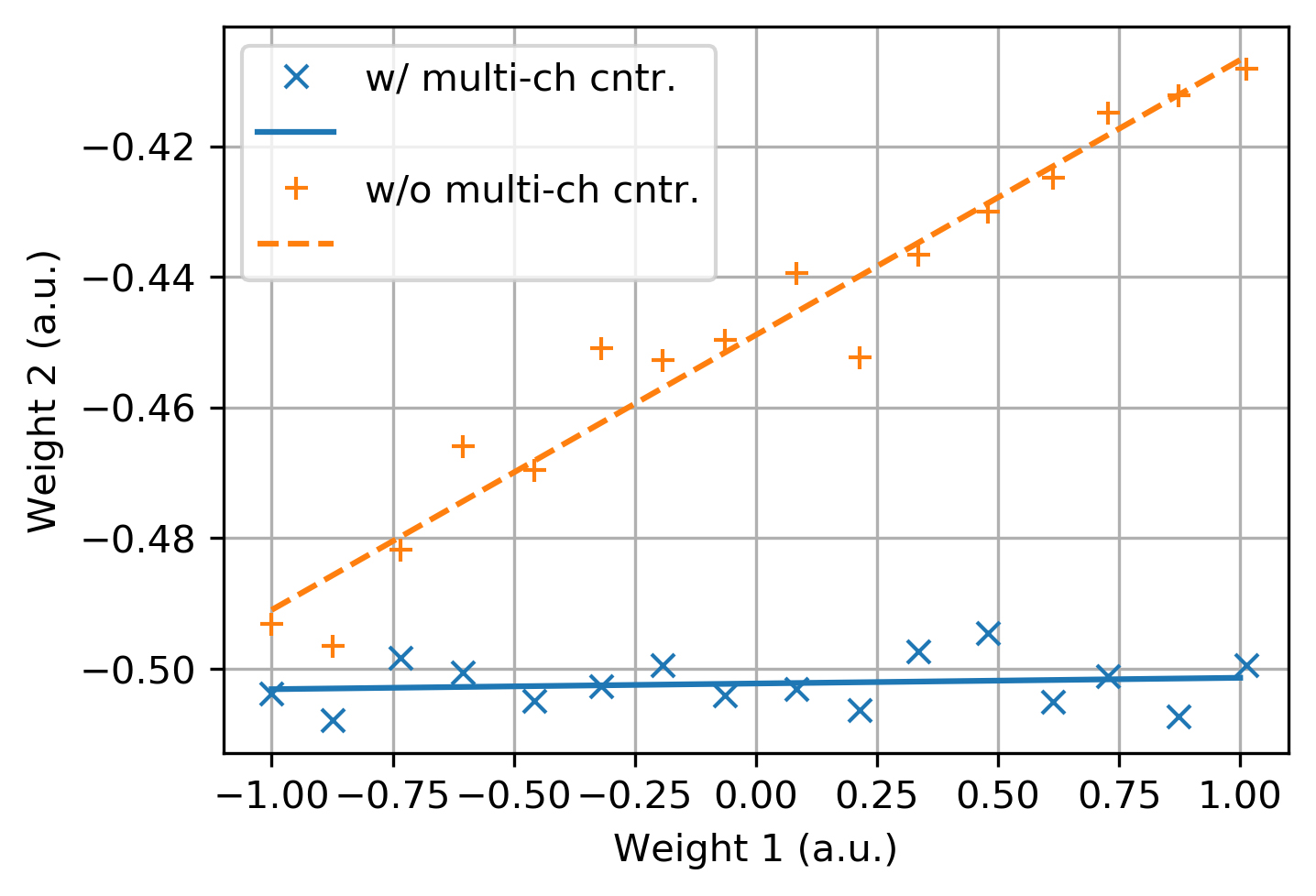}
\caption{Crosstalk suppression using multi-channel dithering control. Solid line: weights of channel 2 obtained by multi-channel control. Dashed line: weights of channel 2 obtained without multi-channel control.
} 
\label{fig:crosstalk}
\end{figure}

Besides the improved accuracy, the multi-dimensional weight searching also alleviates inter-channel crosstalk. Such crosstalk is mainly due to the thermal interference between adjacent photonic devices \cite{jayatilleka2016crosstalk, bogaerts2012silicon}, which causes small weight errors in adjacent MRRs. To evaluate the improvement in crosstalk, we did a comparison test whose results are shown in Figure \ref{fig:crosstalk}. Here, we first tuned the weights to $(w_1,w_2)=(-0.5,-1.0)$, then swept the weight of channel 1 using the dithering weight while keeping a constant current applied to channel 2. This method corresponds to a one-dimensional search in which only channel 1 was considered (refer to the dashed orange curve). Because of the inter-channel crosstalk, when the current applied to MRR 1 is changed, the applied current to MRR 2 needs to be slightly adjusted. Without that adjustment, the weight of MRR 2 drifts by up to 8.8\% ($\max w-\min w = 0.088$), as indicated by the orange curve in the figure. Next, we performed a two-dimensional weight control at the same command weights (solid line in Fig. \ref{fig:crosstalk}). In contrast to the one-dimensional case, the multi-dimensional weight control scheme always considers all channels for every command set of weights. It thus can compensate for crosstalk as indicated by the blue curve, reducing the error to 1.5\% ($\max w-\min w = 0.013$). The nature of the multi-dimensional search keeps this performance advantage valid and more significant as the number of MRR channels scales up.

Another benefit of the multi-dimensional search scheme is avoiding long searching time, especially for large-scale MRR synapses, since it does not scale linearly with the number of MRRs. For example, we compared the time it took to find weight(s) for one MRR and two MRRs simultaneously. After averaging 15 trials, we observed an increase of about $12\%$ for adding a new MRR ($13s$ to $16s$), rather than the expected $100\%$ increase corresponding to a linear increase. This advantage can be explained by the frequency decomposition of the dithering signals: during one search frame, capturing one waveform can effectively measure the weights of all the MRRs. Accordingly, the currents applied to all the MRRs can be simultaneously adjusted in a single step.

\begin{figure}
\centering
\includegraphics[width=.9\linewidth]{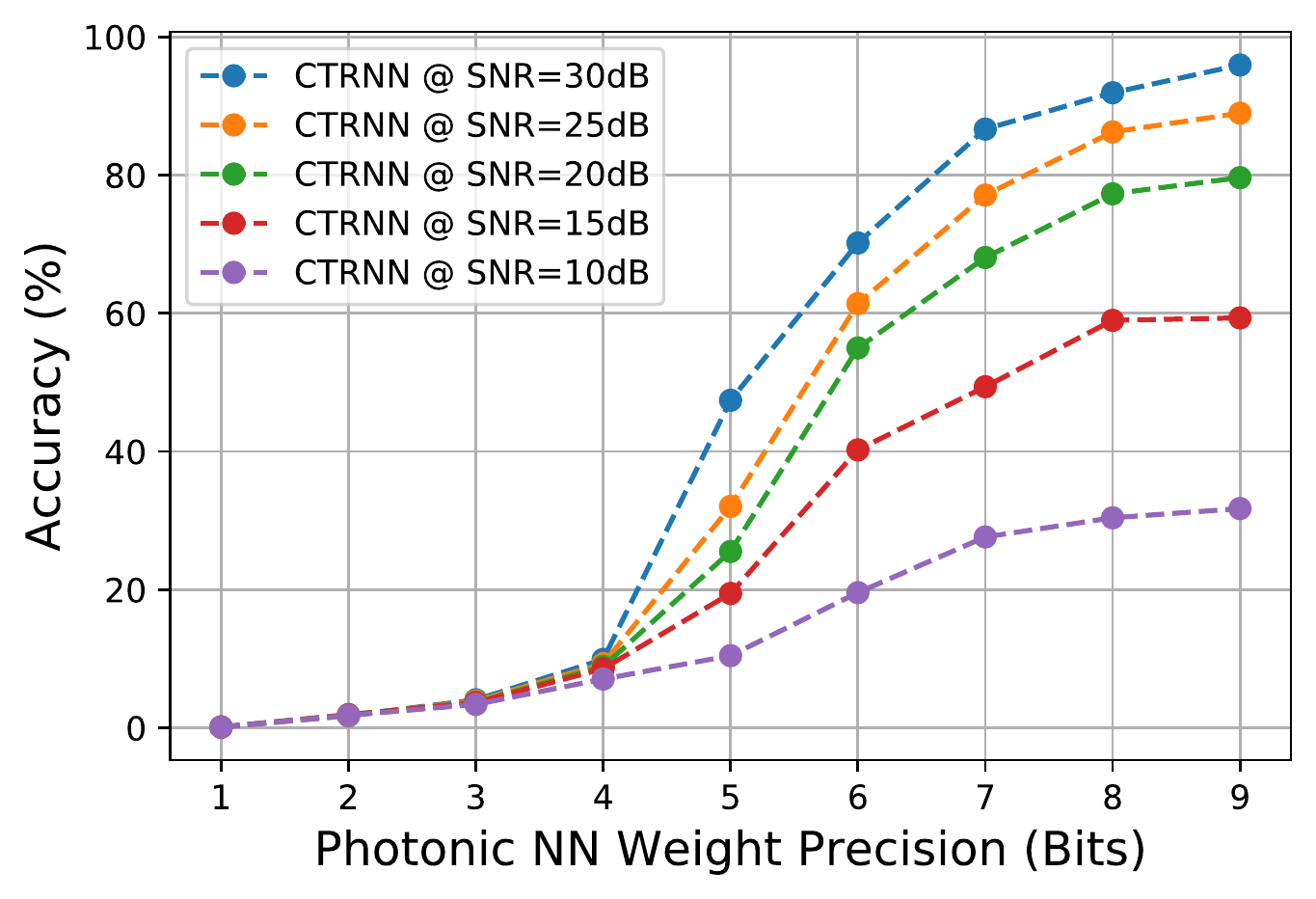}
\caption{Simulation results of prediction accuracy vs bit-precision by a photonic recurrent neural network on a wireless signal classification task under SNR of 10dB, 15dB, 20dB, 25dB and 30dB. 
}
\label{fig:ctrnn}
\end{figure}

To illustrate the impact of the increased bit resolution in photonic neural networks, we performed a numerical simulation of a photonic NN model under varying weight precision levels. We used PyTorch \cite{Paszke2019PyTorchAI} (in Python) to simulate a model of a continuous-time recurrent neural network (CTRNN) \cite{tait2017neuromorphic} (shown in Figure \ref{fig:schematic}(a)), realized by an MRR-based photonic neural network coupled with an FPGA. This network was recently successfully employed for wireless signal classification \cite{peng2021photonic}. The task is to identify 30 identical ZigBee device transmissions by identifying hidden signatures in their RF waveforms. As shown in Figure \ref{fig:ctrnn}, the classification tasks were repeated under different signal-to-noise ratios (SNRs) and weighing precision conditions. As expected, a higher weight precision results in higher prediction accuracy. This is particularly true when inputs have low SNRs since the neural network demands more accurate weighting to separate subtle signatures easily buried under background noise. Assuming a minimum threshold of 80\% accuracy, increasing precision from 7- to 9-bits can compensate for an SNR degradation from 30dB to 20dB. Given this result, this improvement of bit-precision by this proposed dithering control method proves to be practically effective for boosting the performance of photonic NNs, which is especially useful for low-SNR inputs.

The supplemental material includes another numerical simulation on a feedforward DNN, a cascading and fully-connected network with two neurons in each layer. By increasing the network size under the conditions of different bit-precision, the simulation explores how the weighting error (the deviation from the expected weight because of a limited bit-precision) accumulates when expanding the network depth under the conditions of varying bit-precision. We find that increased bit-precision reduces the error accumulated in the network, which allows for large networks. Specifically, we find that a 2-bit improvement in precision (from 7 to 9 bit) enables a network of three times more layers (from 5 to 18) by maintaining the signal-to-noise ratio (SNR=20). This result highlights another benefit of our dithering method: the feasibility of building large networks while maintaining signal fidelity, fitting the needs for more applications.

\section{Conclusion}
In summary, we proposed and tested a dithering control method for photonic synapses realized by MRRs, which can boost the accuracy to 9 bits, which is 2 bits higher than the previous best result. The increased bit precision results in increase of prediction accuracy and enables large networks with a reasonable cost. As aforementioned, our proof-of-principle setup uses a benchtop dithering generator, scope, and CPU to complete the control loop. This limits the speed by a few tens of seconds (for each weight), the precision by 9-bit, the ease in scalability, and the overall footprint. A solution is to use an FPGA with high-speed RF I/O coupled with a lock-in circuit, as shown in \cite{padmaraju2012thermal}, to replace these bulky instruments and carry the dithering control algorithm.

Our dithering approach has several advantages relative to prior work: it is scalable and can accommodate multi-channel MRR synapses, eliminates the crosstalk issue, and speeds up the searching time (thanks to the multi-dimensional searching strategy), requires less frequent calibration, and can be easily incorporated into other photonic synapses. These advantages have great potential for current and future resonator and non-resonator-based techniques, especially in photonic NN applications. The resulting high accuracy makes it competitive against its electronic counterparts, thus benefiting various silicon photonics applications and narrowing the performance gap between electronic and photonic approaches.

\begin{backmatter}
\bmsection{Funding}Defense Advanced Research Projects Agency (HR-00111990049). National Science Foundation (ECCS-1642962). The devices were fabricated at the Advanced Micro Foundry (AMF) in Singapore through the support of CMC Microsystems. B. J. Shastri acknowledges support from the Natural Sciences and Engineering Research Council of Canada (NSERC). S. Bilodeau acknowledges funding from the Fonds de recherche du Québec - Nature et technologies.

\bmsection{Disclosures} The authors declare no conflicts of interest.

\bmsection{Data availability} Data underlying the results presented in this paper are not publicly available at this time but may be obtained from the authors upon reasonable request.

\bmsection{Supplemental document}
See Supplement 1 for supporting content. 

\end{backmatter}

\bibliography{ref}

\bibliographyfullrefs{ref}

\end{document}